# Market-Based Price Autocorrelation

Victor Olkhov
Moscow, Russia
victor.olkhov@gmail.com
ORCID: 0000-0003-0944-5113

## ABSTRACT

This paper assumes that the randomness of market trade values and volumes determines the properties of stochastic market prices. We derive the direct dependence of the first two price statistical moments and price volatility on statistical moments, volatilities, and correlations of market trade values and volumes. That helps describe the dependence of market-based price autocorrelation between times *t* and *t-τ* on statistical moments and correlations between trade values and volumes. That highlights the impact of the randomness of the size of market deals on price statistical moments and autocorrelation. Statistical moments and correlations of market trade values and volumes are assessed by conventional frequency-based probabilities. The distinctions between market-based price autocorrelation and autocorrelation that are assessed by the frequency-based probability analysis of price time series reveal the different approaches to the definitions of price probabilities. To forecast market-based price autocorrelation, one should predict the statistical moments and correlations of trade values and volumes.

This research received no support, specific grant or financial assistance from funding agencies in the public, commercial or nonprofit sectors. We welcome valuable offers of grants, support and positions.

## 1. Introduction

Models of market price autocorrelation help develop forecasts of future price trends. Studies of irregular price time series are based on the use of time series analysis (Woodward, Gray, and Elliott, 1917; Davis, 1941; Cochrane, 2005; Montgomery, Jennings, and Kulahci, 2008). Assessments of price correlations are one of the major ways to uncover internal laws and relations of price evolution and variation.

The description of price autocorrelation is part of a more general problem of the description of time series correlations of economic and financial variables. Economic correlations are discussed in numerous studies (Kendall and Hill, 1953; Friedman, 1962; Fama, 1965; Lo, 1987; Campbell, Grossman, and Wang, 1992; Liu et al., 1997; Plerou et al., 2000; Goetzmann, Li, and Rouwenhorst, 2001; Andersen et al., 2006; Quinn and Voth, 2008; Lind and Ramondo, 2018).

We refer to Fama (1965) and Lo (1987), Liu et al. (1997) and Plerou et al. (2000), Goetzmann, Li, and Rouwenhorst (2001), and Quinn and Voth (2008), who describe different issues of the market price time-series correlations. Karpoff (1987) and Campbell, Grossman, and Wang (1992) model relations between stock market trading volume, price change, and autocorrelations of the daily stock index return.

In our paper, we consider the price autocorrelation as a problem, which is determined by the market origin of price randomness. We emphasize that the assessments of the mean price, price volatility, or price autocorrelation should be determined by the dependence of random market trade of price stochasticity. The time series analysis, regardless of the economic meaning of the particular economic or financial variable, may lead to results with poor economic sense and cause negative, loss-making trade decisions.

In our paper we introduce market-based price autocorrelation and describe their dependence on statistical moments and correlations of market trade values and volumes.

Market time series are very irregular, and any reasonable description of the asset pricing model as well as the description of price correlations should operate with data that is averaged over some time interval $\Delta$. However, market-based price averaging and price autocorrelations can't be evaluated over price time series alone. To describe the impact of random market trade on the statistical properties of market price, one should derive the dependence of price statistical moments and price autocorrelations on statistical moments and correlations of random trade values and volumes. In some extent, this paper develops the results obtained by Olkhov (2021; 2022).

We describe market trade at moment $t_i$ by trade value $C(t_i)$, volume $U(t_i)$, and price $p(t_i)$ that obey trivial trade price equation (1.1):

$$C(t_i) = p(t_i)U(t_i) \qquad (1.1)$$

One can equally describe the properties of a random variable using a probability measure, characteristic function and by a set of statistical moments (Shephard, 1991; Shiryaev, 1999). We consider market trade value $C(t_i)$, volume $U(t_i)$, and price $p(t_i)$ that obey (1.1) as random variables during the averaging interval $\Delta$ and describe them via their statistical moments. It is obvious, that due to (1.1) one can't independently describe statistical moments of random trade value $C(t_i)$, volume $U(t_i)$, and price $p(t_i)$. We call the dependence of price statistical moments and autocorrelations on statistical moments and correlations of random values and volumes as market-based description. In this paper we show how (1.1) generates the dependence of market-based price statistical moments and autocorrelations on statistical moments and correlations of market trade values and volumes.

In section 2, we introduce first two market-based statistical moments of price. In section 4 we derive the dependence of market-based price autocorrelations on statistical moments and correlations of random trade values and volumes. Conclusion in section 4.

We assume that readers are familiar with notions of the probability theory, statistical moments, characteristic function and etc.

## 2. Market-based statistical moments of price

In this section, we introduce the first two market-based statistical moments of price, and in section 3, we show how they help define the market-based price autocorrelation. We refer to Olkhov (2022) for further details. We consider the averaging interval $\Delta$ (2.1) and assume that it contains $N$ terms of random time series of the trade value $C(t_i)$, volume $U(t_i)$, and price $p(t_i)$. We assume that all prices are adjusted to the current time $t$ and that the interval $\varepsilon=t_i-t_{i-1}$ between trades at times $t_i$ and $t_{i-1}$ is constant:

$$\Delta = \left[t - \frac{\Delta}{2}; t + \frac{\Delta}{2}\right] \quad ; \quad t_i \in \Delta \quad ; \quad i = 1, \dots N \qquad (2.1)$$

The standard interpretation of the price $p$ probability "is based on the probabilistic approach and using A. N. Kolmogorov's axiomatic probability theory, which is generally accepted now" (Shiryaev, 1999). If there are $N$ trades during the interval $\Delta$ (2.1) and there are $n(p)$ trades at price $p$, then the ratio $n(p)/N$ gives the assessment of the probability $P(p)$ of price $p$. The frequency $n(p)/N$ of events is the conventional assessment of probability, and we call it the frequency-based approach to price probability to distinguish it from the market-based approach. The assessments of the frequency-based *n-th* statistical moments of a random variable $q_i$ , $i=1,..N,$ by the finite number $N$ of terms that belong to the averaging interval $\Delta$ have the form:

$$E[q_i^n] \sim \frac{1}{N} \sum_{i=1}^{N} q_i^n \tag{2.2}$$

We denote $E[..]$ mathematical expectation of the frequency-based probability and use ~ to highlight that (2.2) gives the assessment of the *n-th* statistical moment (2.2) of a random variable $q$ by the finite $N$ of terms.

### *2.1 Preliminary assessments*

The frequency-based probabilities (2.2) define the *n-th* statistical moments of the market trade value $C(t;n)$ and volume $U(t;n)$ during the interval $\Delta$:

$$C(t;n) = E[C^n(t_i)] \sim \frac{1}{N} \sum_{i=1}^{N} C^n(t_i) \quad ; \quad U(t;n) = E[U^n(t_i)] \sim \frac{1}{N} \sum_{i=1}^{N} U^n(t_i) \tag{2.3}$$

The frequency-based assessment of the *n-th* statistical moments of price $p(t;n)$ takes the form:

$$p(t;n) = E[p^n(t_i)] \sim \frac{1}{N} \sum_{i=1}^{N} p^n(t_i) \tag{2.4}$$

It is obvious that the equation (1.1) prohibits the usage of independent definitions of trade value $C(t;n),$ volume $U(t;n)$ (2.3), and price $p(t;n)$ (2.4) statistical moments. Actually, (2.4) is not the only description of price statistical moments. Volume weighted average price (VWAP) was introduced by Berkowitz et al. (1988) and is widely in use now (Buryak and Guo, 2014; Busseti and Boyd, 2015). We denote VWAP as $p(t;1,1)$ :

$$p(t;1,1) = \frac{\sum_{i=1}^{N} p(t_i)U(t_i)}{\sum_{i=1}^{N} U(t_i)} = \sum_{i=1}^{N} p(t_i)w(t_i;1) = \frac{C_\Sigma(t;1)}{U_\Sigma(t;1)} = \frac{C(t;1)}{U(t;1)} \tag{2.5}$$

$$C(t;1) = p(t;1,1)U(t;1) \quad ; \quad C_\Sigma(t;1) = p(t;1,1)\, U_\Sigma(t;1) \tag{2.6}$$

$$C_\Sigma(t;1) = \sum_{i=1}^{N} C(t_i) \quad ; \quad U_\Sigma(t;1) = \sum_{i=1}^{N} U(t_i) \tag{2.7}$$

$$w(t_i;1) = \frac{U(t_i)}{U_\Sigma(t;1)} \quad ; \quad \sum_{i=1}^{N} w(t_i;1) = 1 \tag{2.8}$$

Relations (2.5; 2.6) describe VWAP $p(t;1,1),$ and equations (2.6) reproduce the form of the trade price equation (1.1). Relations (2.7) define the total value $C_\Sigma(t;1)$ and the total volume $U_\Sigma(t;1)$ of market trades during the averaging interval $\Delta$ (2.1). The weight functions $w(t_i;1)$ (2.8) determine VWAP (2.5) and highlight the distinction between VWAP (2.5) and the frequency-based average price $p(t;1)$ (2.4). The frequency-based price probability (2.2) proposes that all trades during the interval $\Delta$ (2.1) have equal weights $\sim 1/N$. Contrary to that, VWAP uses the assumption that each trade at time $t_i$ has a weight $w(t_i;1)$ (2.8) proportional to the size of the trade volume $U(t_i)$. If all trade volumes $U(t_i)$ are *constant* during $\Delta$ (2.1), then VWAP $p(t;1,1)$ (2.5) coincides with the conventional frequency-based average price $p(t;1)$ (2.4). The relations (2.5-2.8) describe the impact of the random trade values $U(t_i)$ on VWAP $p(t;1,1)$. Actually, the trade price equation generates a set of averaging relations similar to (2.5-2.8). Let us take the *m-th* degree of (1.1):

$$C^m(t_i) = p^m(t_i)U^m(t_i) \quad ; \quad m = 1,2,3,\ldots \tag{2.9}$$

The equations (2.9) generate a set of averaging relations similar to (2.5-2.8):

$$p(t;n,m) = \sum_{i=1}^{N} p^n(t_i)w(t_i;m) = \frac{1}{\sum_{i=1}^{N} U^m(t_i)} \sum_{i=1}^{N} p^n(t_i)U^m(t_i) \tag{2.10}$$

$$w(t_i;m) = \frac{U^m(t_i)}{\sum_{i=1}^{N} U^m(t_i)} \quad ; \quad \sum_{i=1}^{N} w(t_i;m) = 1 \tag{2.11}$$

For each $m=1,2,..$, (2.10) describes the averaging of the *n-th* degree of price $p^n(t_i)$ via the *m-th* weight functions $w(t_i;m)$ (2.11) that correspond to the *m-th* equation (2.9). If $n=m$ then relations (2.10) take the form (2.12-2.14) alike to (2.5):

$$p(t;n,n) = \frac{1}{\sum_{i=1}^{N} U^n(t_i)} \sum_{i=1}^{N} p^n(t_i)U^n(t_i) = \frac{C_\Sigma(t;n)}{U_\Sigma(t;n)} = \frac{C(t;n)}{U(t;n)} \tag{2.12}$$

$$C(t;n) = p(t;n,n)U(t;n) \quad ; \quad C_\Sigma(t;n) = p(t;n,n)\, U_\Sigma(t;n) \tag{2.13}$$

$$C_\Sigma(t;n) = \sum_{i=1}^{N} C^n(t_i) \quad ; \quad U_\Sigma(t;n) = \sum_{i=1}^{N} U^n(t_i) \tag{2.14}$$

The set of equations (2.9) for each $m=1,2,3,..$ generates the *n-th* statistical moments of price $p(t;n,m)$ that are determined by the weight functions $w(t_i;m)$ (2.11). If all trade volumes $U(t_i)$ are constant during $\Delta$ (2.1), then all weight functions $w(t_i;m)=1/N$ and (2.10) coincide with the conventional frequency-based statistical moments of price $p(t;n)$ (2.4).

### 2.2 Market-based statistical moments

Relations (2.5-2.14) form the basis for definition of market-based statistical moments of price. To highlight the distinction between the market-based price probability and the conventional frequency-based probability, we shall note market-based mathematical expectation as $E_m[..]$.

We propose to use VWAP $p(t;1,1)$ (2.5) as the market-based average price $E_m[p(t_i)]=a(t;1)$ and suggest one more argument in favor of that choice. Indeed, all investors assess the average price of the shares in their portfolio by the ratio of the total expenses they spent to purchase the shares to the total number of shares. Exactly the same relations are described by the ratio (2.5) of the total value $C_\Sigma(t;1)$ to the total volume $U_\Sigma(t;1)$ (2.7) of market trades during the averaging interval $\Delta$ (2.1). The choice of market-based average price $a(t;1)$:

$$E_m[p(t_i)] = a(t;1) = p(t;1,1) \tag{2.15}$$

impacts the definitions of the 2-d market-based statistical moment $a(t;2)$:

$$E_m[p^2(t_i)] = a(t;2) \tag{2.16}$$

Nevertheless, relations (2.5-2.14) present a wide choice for the 2-d market-based statistical moment $a(t;2)$, such a choice should be consistent with the market-based average price $a(t;1)$. In particular, market-based price volatility $\sigma^2(t)$ (2.17) that is determined by the first two statistical moments $a(t;1)$ and $a(t;2)$ should be non-negative:

$$\sigma^2(t) = E_m[(p(t_i) - a(t;1))^2] = a(t;2) - a^2(t;1) \geq 0 \tag{2.17}$$

To fulfill that condition, we take the $\sigma^2(t)$ (2.17) to be equal (2.18):

$$\sigma^2(t) = E_m[(p(t_i) - a(t;1))^2] = \sum_{i=1}^{N}(p(t_i) - a(t;1))^2 w(t_i;2) = M(t) \tag{2.18}$$

Equation (2.18) ties up price volatility $\sigma^2(t)$ (2.17) or second central market-based statistical moment with the mean square of the fluctuations of price near its average $a(t;1)$ averaged via the weight functions $w(t_i;2)$ (2.11). From (2.10) obtain:

$$M(t) = p(t;2,2) - 2a(t;1)p(t;1,2) + a^2(t;1) = \sigma^2(t) \tag{2.19}$$

Equation (2.19) defines the market based 2-d statistical moment $a(t;2)$ that always fulfills the inequality (2.17):

$$a(t;2) = p(t;2,2) - 2a(t;1)p(t;1,2) + 2a^2(t;1) \tag{2.20}$$

Let us highlight the dependence of first two market-based statistical moments $a(t;1)$ (2.15) and $a(t;2)$ (2.20) on statistical moments and correlations of the trade values and volumes.

$$a(t;1) = \frac{C(t;1)}{U(t;1)} \quad ; \quad p(t;2,2) = \frac{C(t;2)}{U(t;2)} \tag{2.21}$$

$$p(t;1,2) = \frac{1}{U(t;2)}\frac{1}{N}\sum_{i=1}^{N} p(t_i)U^2(t_i) = \frac{1}{U(t;2)}\frac{1}{N}\sum_{i=1}^{N} C(t_i)U(t_i) = \frac{CU(t;1)}{U(t;2)} \tag{2.22}$$

As $CU(t;1)$ we denote the joint frequency-based statistical moment (2.23) of the trade value $C(t_i)$ and volume $U(t_i)$:

$$CU(t;1) = E[C(t_i)U(t_i)] \sim \frac{1}{N}\sum_{i=1}^{N} C(t_i)U(t_i) = C(t;1)U(t;1) + corr\{C(t)U(t)\} \tag{2.23}$$

From (2.20; 2.22; 2.23) obtain:

$$\sigma^2(t) = \frac{\Omega_C^2(t) + a^2(t;1)\Omega_U^2(t) - 2a(t;1)corr\{C(t)U(t)\}}{U(t;2)} \tag{2.24}$$

$$a(t;2) = \frac{C(t;2) + 2a^2(t;1)\Omega_U^2(t) - 2a(t;1)corr\{C(t)U(t)\}}{U(t;2)} \tag{2.25}$$

As $\Omega_C^2(t)$ and $\Omega_U^2(t)$ (2.26) we denote volatilities of market trade values and volumes:

$$\Omega_C^2(t) = C(t;2) - C^2(t;1) \quad ; \quad \Omega_U^2(t) = U(t;2) - U^2(t;1) \tag{2.26}$$

Relations (2.21; 2.24; 2.25) highlight the direct dependence of market-based statistical moments of price $a(t;1)$, $a(t;2)$, and price volatility $\sigma^2(t)$ on statistical moments, volatilities, and correlations of market trade values and volumes. Predictions of market-based price volatility $\sigma^2(t)$ (2.24) depend on the forecasts of the statistical moments and correlations of market trade values and volumes.

## 3. Market-based price autocorrelations

In this section, we apply the above method that defines the 2-d market-based price statistical moment $a(t;2)$ and the volatility $\sigma^2(t)$ to derive the dependence of the market-based autocorrelation $corr\{p(t)p(t-\tau)\}$ on statistical moments and correlations of the trade values and volumes.

Let us consider the trade price equation (1.1) at times $t_i$ and $t_i$-$\tau$ and take their product:

$$C(t_i)C(t_i-\tau) = p(t_i)p(t_i-\tau)U(t_i)U(t_i-\tau) \tag{3.1}$$

We assume that $t_i$ belong to the averaging interval $\Delta$ (2.1) and time shift $\tau= l\cdot\varepsilon$, $l=1,2,\ldots$, is multiple of $\varepsilon$, so $t_i$-$\tau$ corresponds to particular trade in the past. Similar to (2.3) we define the joint statistical moments $C(t,\tau)=E[C(t_i)C(t_i-\tau)]$ of trade values and trade volumes $U(t,\tau)=E[U(t_i)U(t_i-\tau)]$:

$$C(t,\tau) = E[C(t_i)C(t_i-\tau)] \sim \frac{1}{N}\sum_{i=1}^{N} C(t_i)C(t_i-\tau) \;\; ; \; C_\Sigma(t,\tau) = N\cdot C(t,\tau) \tag{3.2}$$

$$U(t,\tau) = E[U(t_i)U(t_i-\tau)] \sim \frac{1}{N}\sum_{i=1}^{N} U(t_i)U(t_i-\tau) \;\; ; \; U_\Sigma(t,\tau) = N\cdot U(t,\tau) \tag{3.3}$$

The relations (3.2; 3.3) and (2.3) define autocorrelations $corr\{C(t)C(t-\tau)\}$ of trade values and $corr\{U(t)U(t-\tau)\}$ trade volumes:

$$C(t,\tau) = C(t;1)C(t-\tau) + corr\{C(t)C(t-\tau)\} \tag{3.4}$$

$$U(t,\tau) = U(t;1)U(t-\tau) + corr\{U(t)U(t-\tau)\} \tag{3.5}$$

The equation (3.1) has the form similar to (1.1) and alike to (2.8; 2.11) we define the weight functions $w(t,\tau)$ that evaluate averaging for (3.1):

$$w(t_i,\tau) = \frac{U(t_i)U(t_i-\tau)}{U_\Sigma(t,\tau)} \quad ; \quad \sum_{i=1}^{N} w(t_i,\tau) = 1 \tag{3.6}$$

Similar to (2.5; 2.10), the joint statistical moments $C(t,\tau)$ (3.2) of trade values and $U(t,\tau)$ of trade volumes, and the weight functions (3.6) define the joint price expectation $p(t,\tau)$:

$$p(t,\tau) = \sum_{i=1}^{N} p(t_i)p(t_i-\tau)w(t_i,\tau) = \frac{1}{U_\Sigma(t,\tau)}\sum_{i=1}^{N} p(t_i)p(t_i-\tau)U(t_i)U(t_i-\tau) \tag{3.7}$$

From (3.1-3.3) obtain:

$$p(t,\tau) = \frac{C_\Sigma(t,\tau)}{U_\Sigma(t,\tau)} = \frac{C(t,\tau)}{U(t,\tau)} \tag{3.8}$$

We define the market-based joint price statistical moment $a(t,\tau)$ (3.9):

$$a(t,\tau) = E_m[p(t_i)p(t_i-\tau)] = E_m[p(t_i)]E_m[p(t_i-\tau)] + corr\{p(t)p(t-\tau)\}$$

$$a(t,\tau) = a(t;1)a(t-\tau) + corr\{p(t)p(t-\tau)\} \tag{3.9}$$

The choice of $a(t,\tau)$ (3.9) should be consistent with market based 1-st statistical moments $a(t;1)$ and $a(t$-$\tau;1)$ (2.15), and price autocorrelation $corr\{p(t)p(t-\tau)\}$ should be consistent with price volatility $\sigma^2(t)$ (2.17). For $\tau=0$ price autocorrelation $corr\{p(t)p(t-\tau)\}$ should be equal to the price volatility $\sigma^2(t)$ (2.17):

$$corr\{p(t)p(t-\tau)\}|_{\tau=0} = \sigma^2(t) = a(t;2) - a^2(t;1) \quad (3.10)$$

To fulfill these conditions, we consider the product of price fluctuations near their average values

$$\delta p(t_i,\tau) = [p(t_i) - a(t;1)][p(t_i-\tau) - a(t-\tau;1)] \quad (3.11)$$

and define $\delta p(t,\tau)$ (3.12) as the average of $\delta p(t_i,\tau)$ (3.11) via the weight functions $w(t_i,\tau)$ (3.6):

$$\delta p(t,\tau) = \sum_{i=1}^{N} \delta p(t_i,\tau) w(t_i,\tau) \quad (3.12)$$

To fulfill the condition (3.10), we take the price autocorrelation $corr\{p(t)p(t-\tau)\}$ to be equal to $\delta p(t,\tau)$ (3.12). In Appendix A, we present simple but lengthy calculations that derive the dependence of $a(t,\tau)$ (3.9) and the price autocorrelation $corr\{p(t)p(t-\tau)\}$ (3.9) on statistical moments and correlations of market trade values and volumes. The market-based price autocorrelation is determined by (A.15) and takes the form:

$$corr\{p(t)p(t-\tau)\} = \frac{corr\{C(t)C(t-\tau)\} + a(t;1)a(t-\tau;1)corr\{U(t)U(t-\tau)\}}{U(t,\tau)} -$$

$$- \frac{a(t-\tau;1)corr\{C(t)U(t-\tau)\} + a(t;1)corr\{C(t-\tau)U(t)\}}{U(t,\tau)} \quad (3.13)$$

The market-based joint price expectation $a(t,\tau)$ (3.9) takes the form (A.16):

$$a(t,\tau) = \frac{C(t,\tau) + 2a(t;1)a(t-\tau;1)corr\{U(t)U(t-\tau)\}}{U(t,\tau)} -$$

$$- \frac{a(t-\tau;1)corr\{C(t)U(t-\tau)\} + a(t;1)corr\{C(t-\tau)U(t)\}}{U(t,\tau)} \quad (3.14)$$

In App.A., we show that for $\tau=0$, the price autocorrelation (3.13) equals the price volatility $\sigma^2(t)$ (2.24) and the joint price expectation $a(t,\tau)$ (3.14) equals the 2-d market-based price statistical moment $a(t;2)$ (2.25). That confirms the correctness of the derivation of the price autocorrelation.

## 4. Conclusion

Market-based price statistical moments $a(t;1)$ (2.15) and $a(t;2)$ (2.16; 2.25), price volatility $\sigma^2(t)$ (2.24), and price autocorrelation (3.13) differ from the frequency-based assessments (Kendall and Hill, 1953; Fama, 1965; Lo, 1987; Liu et al., 1997; Plerou et al., 2000; Goetzmann, Li, and Rouwenhorst, 2001; Quinn and Voth, 2008). That is the result of the different assessments of price randomness. As usual, researchers consider the random price time series as the only source for frequency-based assessments of random price properties. Contrary to that, we highlight that, due to the primitive trade price equation (1.1), the stochasticity of market trade values and volumes should determine the properties of a random price. We derive the dependence of statistical moments and correlations of trade values and volumes on market-based statistical moments and the autocorrelation of random prices.

Such a market-based approach to market price statistics emphasizes its dependence on the random properties of market trades. That makes the forecasts of average price, price volatility, and correlations independently from the predictions of the statistical moments and correlations of market trades not too reliable. It is obvious that the unified description is too complex, and different approximations should be developed. However, each particular approximation should maintain the economic meaning of the variables. The conventional frequency-based approach to random price helps study price time series as self-sufficient elements, simplifies the problem, but the results may have poor economic meaning and do not support the dependence on properties of random market trades.

However, economics is a social science and highly depends on agents' views, habits, expectations, and preferences. Economic processes and market transactions in particular, are mostly governed by agents' decisions. Agents are free in their choice of preferred definitions of price statistical moments and price autocorrelation, no matter what economic sense they have. Agents may base their trade decisions on conventional standard frequency-based price probability and price statistical moments (2.4) or follow market-based price statistical moments and autocorrelation (2.15; 2.25; 3.13). Agents' habits and beliefs may make the use of frequency-based price probability more preferred. However, a permanent mismatch between frequency-based assessments of price statistical moments and real market outcomes may add scores in favor of the market-based description of price statistical moments and price autocorrelation in particular. Actually, it will be more difficult, but the days of simple solutions have long gone.

The choice of market-based price statistical moments and price autocorrelation may require reassessments of the observed market time series and correction of conventional conclusions. That may deliver a new vision of market price statistical properties that should more rely on random market trade behavior. It is clear that proposed transition from frequency-based to market-based price statistical moments and price autocorrelation will not simplify the problem. However, it may help develop approximations that will give more adequate predictions of market reality.

## Appendix A

## Derivation of market-based price autocorrelation

To fulfill the conditions (3.9; 3.10) we state that the price autocorrelation *corr{p(t)p(t-τ)}* (3.9) should be equal to the $\delta p(t,\tau)$ (3.12), which denotes the averaging of $\delta p(t_i,\tau)$ (3.11) via the weight functions $w(t,\tau)$ (3.6):

$$corr\{p(t)p(t-\tau)\} = \delta p(t,\tau) = \sum_{i=1}^{N} \delta p(t_i,\tau) w(t_i,\tau) \quad (A.1)$$

To evaluate the right-hand side in (A.1), we use (3.11):

$$\delta p(t_i,\tau) = p(t_i)p(t_i-\tau) - a(t;1)p(t_i-\tau) - a(t-\tau;1)p(t_i) + a(t;1)a(t-\tau;1) \quad (A.2)$$

Now we average each term of (A.2) via the weight functions (3.6). From (3.7; 3.8), obtain:

$$p(t,\tau) = \sum_{i=1}^{N} p(t_i)p(t_i-\tau)w(t_i,\tau) = \frac{C(t,\tau)}{U(t,\tau)} \quad (A.3)$$

$$p(t,\tau;1) = \sum_{i=1}^{N} p(t_i)w(t_i,\tau) = \frac{1}{N\cdot U(t,\tau)}\sum_{i=1}^{N} p(t_i)U(t_i)U(t_i-\tau) \quad (A.4)$$

From the trade price equation (1.1), obtain:

$$p(t,\tau;1) = \frac{\sum_{i=1}^{N} C(t_i)U(t_i-\tau)}{N\cdot U(t,\tau)} = \frac{CU(t,t-\tau)}{U(t,\tau)} \quad (A.5)$$

We denote *CU(t,t-τ)* as the joint expectation (A.6) of trade value $C(t_i)$ and volume $U(t_i-\tau)$:

$$CU(t,t-\tau) = E[C(t_i)U(t_i-\tau)] = \frac{1}{N}\sum_{i=1}^{N} C(t_i)U(t_i-\tau) \quad (A.6)$$

The joint expectation *CU(t,t-τ)* determines the correlation *corr{C(t)U(t-τ)}:*

$$CU(t,t-\tau) = E[C(t_i)]E[U(t_i-\tau)] + corr\{C(t)U(t-\tau)\}$$

$$CU(t,t-\tau) = C(t;1)U(t-\tau;1) + corr\{C(t)U(t-\tau)\} \quad (A.7)$$

Similar to (A.4; A.5), we derive $p(t,\tau;2)$:

$$p(t,\tau;2) = \sum_{i=1}^{N} p(t_i-\tau)w(t_i,\tau) = \frac{1}{N\cdot U(t,\tau)}\sum_{i=1}^{N} p(t_i-\tau)U(t_i)U(t_i-\tau) \quad (A.8)$$

$$p(t,\tau;2) = \frac{\sum_{i=1}^{N} C(t_i-\tau)U(t_i)}{N\cdot U(t,\tau)} = \frac{CU(t-\tau,t)}{U(t,\tau)} \quad (A.9)$$

The joint expectation *CU(t-τ,t)* (A.10) determines the correlation *corr{C(t-τ)U(t)}* (A.11)

$$CU(t-\tau,t) = E[C(t_i-\tau)U(t_i)] = \frac{1}{N}\sum_{i=1}^{N} C(t_i-\tau)\, U(t_i) \quad (A.10)$$

$$CU(t-\tau,t) = C(t-\tau;1)U(t;1) + corr\{C(t)U(t-\tau)\} \quad (A.11)$$

Finally, from (A.2-A.12), we derive the expression for price autocorrelation *corr{p(t)p(t-τ)}*:

$$corr\{p(t)p(t-\tau)\} = p(t,\tau) - a(t;1)p(t,\tau;2) - a(t-\tau;1)p(t,\tau;1) + a(t;1)a(t-\tau;1) \quad (A.12)$$

With the help of (A.3; A.5; A.8) one can transform (A.12) as the follows:

$$corr\{p(t)p(t-\tau)\} = \frac{C(t,\tau) - a(t;1)CU(t,t-\tau) - a(t-\tau;1)CU(t,t-\tau)}{U(t,\tau)} + a(t;1)a(t-\tau;1) \quad (A.13)$$

The relations (A.7; A.10) permit transfer (A.13) as follows:

$$corr\{p(t)p(t-\tau)\} =$$

$$= \frac{C(t;1)C(t-\tau) - a(t-\tau;1)C(t;1)U(t-\tau;1) - a(t;1)C(t-\tau;1)U(t;1)}{U(t,\tau)} -$$

$$- \frac{a(t-\tau;1)corr\{C(t)U(t-\tau)\} + a(t;1)corr\{C(t-\tau)U(t)\} - corr\{C(t)C(t-\tau)\}}{U(t,\tau)} + a(t;1)a(t-\tau;1) \quad (A.14)$$

The use of (2.21) that ties up the average value *C(t;1),* volume *U(t;1),* and price *a(t;1):*

$$C(t;1) = a(t;1)U(t;1)$$

gives the final expression for the price autocorrelation:

$$corr\{p(t)p(t-\tau)\} = \frac{corr\{C(t)C(t-\tau)\} + a(t;1)a(t-\tau;1)corr\{U(t)U(t-\tau)\}}{U(t,\tau)} -$$

$$- \frac{a(t-\tau;1)corr\{C(t)U(t-\tau)\} + a(t;1)corr\{C(t-\tau)U(t)\}}{U(t,\tau)} \quad (A.15)$$

Similar calculations that take into account (3.9) give the expression of *a(t,τ):*

$$a(t,\tau) = \frac{C(t,\tau) + 2a(t;1)a(t-\tau;1)corr\{U(t)U(t-\tau)\} - a(t-\tau;1)corr\{C(t)U(t-\tau)\} - a(t;1)corr\{C(t-\tau)U(t)\}}{U(t,\tau)} \quad (A.16)$$

One can check that for *τ=0,* the price autocorrelation (A.15) equals the price volatility $\sigma^2(t)$:

$$corr\{C(t)C(t-\tau)\}|_{\tau=0} = C(t;2) - C^2(t;1) = \Omega_C^2(t)$$

$$a(t;1)a(t-\tau;1)corr\{U(t)U(t-\tau)\}|_{\tau=0} = a^2(t;1)\Omega_U^2(t)$$

$$a(t-\tau;1)corr\{C(t)U(t-\tau)\} + a(t;1)corr\{C(t-\tau)U(t)\}|_{\tau=0} = 2a(t;1)corr\{C(t)U(t)\}$$

$$U(t,\tau)|_{\tau=0} = U(t;2)$$

$$corr\{p(t)p(t-\tau)\}|_{\tau=0} = \frac{\Omega_C^2(t) + a^2(t;1)\Omega_U^2(t) - 2a(t;1)corr\{C(t)U(t)\}}{U(t;2)} = \sigma^2(t)$$

One obtains that for *τ=0,* the price autocorrelation (A.15) coincides with the price volatility $\sigma^2(t)$ (2.24). The similar relations prove that for *τ=0, a(t,τ)* (A.16) equals *a(t;2)* (2.25).

That confirms the correct choice of the joint price statistical moment *a(t,τ)* (A.16) and price autocorrelation (A.15) and their consistency with the first two market-based price statistical moments *a(t;1)* (2.15) and *a(t;2)* (2.25).